\documentclass[review,authoryear,12pt]{elsarticle}

\usepackage{graphicx}
\usepackage[tight,footnotesize, nooneline]{subfigure}

\usepackage{adjustbox} 
\usepackage[hidelinks]{hyperref} 

\usepackage{amssymb}
\usepackage{amsthm}
\usepackage{amsmath}

\usepackage{lineno}
 
\usepackage{multirow}

\usepackage[flushleft]{threeparttable} 

\journal{arXiv}

\newcommand\copyrighttext{%
   \textcopyright\ 2019 IEEE.}

\begin{document}

\begin{frontmatter}



\title{High-intensity focused ultrasound therapy in the uterine fibroid: a clinical case study of poor heating efficacy}


\author[Affil1]{Visa Suomi \corref{cor1}}
\author[Affil1]{Antti Viitala}
\author[Affil2]{Teija Sainio}
\author[Affil1]{Gaber Komar}
\author[Affil1]{Roberto Blanco Sequeiros}

\address[Affil1]{Department of Radiology, Turku University Hospital, Kiinamyllynkatu 4-8, 20521 Turku, Finland}
\address[Affil2]{Department of Medical Physics, Turku University Hospital, Kiinamyllynkatu 4-8, 20521 Turku, Finland}
\cortext[cor1]{Corresponding Author: Visa Suomi, Department of Radiology, Turku University Hospital, Kiinamyllynkatu 4-8, 20521 Turku, Finland; Email, visa.suomi@tyks.fi}

\begin{abstract}
A clinical case study of high-intensity focused ultrasound (HIFU) treatment in the uterine fibroid was conducted. During the therapy, poor heating efficacy was observed which could be attributed to several factors such as the local perfusion rate, patient-specific anatomy or changes in acoustic parameters of the ultrasound field. In order to determine the cause of the diminished heating, perfusion analyses and ultrasound simulations were conducted using the magnetic resonance imaging (MRI) data from the treatment. The perfusion analysis showed high local perfusion rate in the myoma (301.0 $\pm$ 25.6~mL/100 g/min) compared to the surrounding myometrium (233.8 $\pm$ 16.2~mL/100 g/min). The ultrasound simulations did not show large differences in the focal point shape or the acoustic pressure (2.07 $\pm$ 0.06~MPa) when tilting the transducer. However, a small shift ($-$2.2 $\pm$ 1.3~mm) in the axial location of the focal point was observed. The main causes for the diminished heating were likely the high local perfusion and ultrasound attenuation due to the deep location of the myoma.
\end{abstract}


\end{frontmatter}

\copyrighttext

\pagebreak



\section*{Introduction}

High-intensity focused ultrasound (HIFU) therapy is a non-invasive treatment method which can be used to treat uterine fibroids (i.e., myomas)~\citep{kim2011mr}. The clinical treatment of patients with uterine fibroids using HIFU therapy is nowadays a standard practice, but the outcome of the treatment is sometimes unsuccessful due to the diminished heating efficacy in the target region~\citep{thiburce2015magnetic}. The lower heating effect in some patients can be attributed to several factors such as the local perfusion rate, the location and depth of the myoma, and the attenuation of ultrasound energy~\citep{rueff2013clinical, kim2014techniques, wang2016influence, suomi2018full}. Furthermore, in some cases the transducer has to be positioned at an angle in order to avoid intervening tissue structures or to reach deep-laying targets~\citep{sainio2018wedged}. All of these effects reduce the heating efficacy resulting in a poor response to the HIFU treatment.

The aim of this study was to investigate how local perfusion, transducer positioning and patient-specific anatomy affect the treatment efficacy. For this purpose, a retrospective clinical case study of HIFU treatment in the uterine fibroid with diminished heating was conducted. Magnetic resonance imaging (MRI) was used to determine the patient anatomy as well as to quantify the local perfusion rate of the myoma. Furthermore, three-dimensional ultrasound simulations were conducted on the segmented MR-image data in order to evaluate the effect of transducer orientation and patient anatomy on the acoustic parameters of the ultrasound focal point. The results identify the main factors affecting the treatment outcome, which helps avoiding similar cases in the future.

\section*{Clinical case}

\begin{figure}[b!]
    \centering
    \subfigure[]
    {
        \includegraphics[width=0.45\columnwidth]{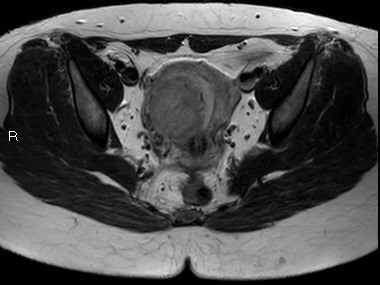}
    }
    \subfigure[]
    {
        \includegraphics[width=0.45\columnwidth]{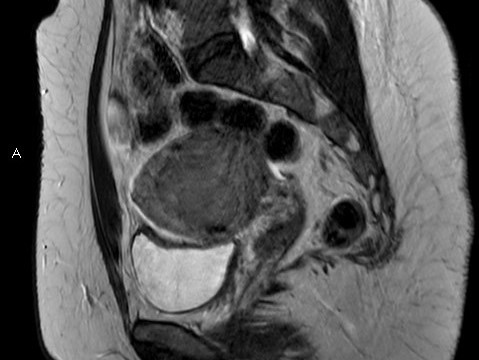}
    }
    \caption{(a) Axial and (b) sagittal T2w-MR images of the myoma from the pre-treatment screening scan. The fibroid appeared hypointense when compared to the myometrium.}
    \label{fig:patient_geometry}
\end{figure}

The study was retrospective using the patient data from a uterine fibroid HIFU treatment conducted at the Turku University Hospital, Finland. Ethical permission for the usage of the patient data (ETMK: 152/1801/2016) was obtained from the Ethics Committee of Hospital District of Southwest Finland. All patient records collected for the study were anonymised prior using the data. The experimental procedures involving human subjects described in this paper were approved by the Institutional Review Board.

The patient was 40 years old with no history of pregnancies or previous uterine fibroid treatments. 
She had been diagnosed with a single 59~mL intramuscular leiomyoma located in the lateral myometrium. 
In the pre-treatment T2w-MRI scan (see Fig.~\ref{fig:patient_geometry}), the fibroid was hypointense when compared to the myometrium with the signal intensity values of 66.6, 224.9 and 175.6 for abdominal muscle, myometrium and myoma, respectively. In the gadolinium-enhanced contrast scan, the fibroid was markedly hyperintense with respect to myometrium during the arterial phase, but hypointense during the venous phase. In addition to MRI examination, partial resection was performed to obtain tissue sample and confirm the diagnosis of benign uterine fibroid. During the resection, tissue was also found to be highly vascularised.

The patient was treated using a clinical HIFU therapy system (Sonalleve V2, Profound Medical, Mississauga, Canada). During the HIFU therapy, adequate heating was not observed (the highest temperature was 55.4~$^{\circ}$C) despite using 300~W of acoustic power which is the maximum allowed by the therapy system. Factors that could potentially contribute to the poor heating during the therapy include: (i) high perfusion rate of the tissue, (ii) the rotation of the transducer during the sonications, and (iii) the use of wedge-shaped gel pad to manipulate the patient anatomy.

In the post-therapy MRI scans, decent non-perfused volume (NPV) ratio (i.e., the ratio of the ablated volume to that of the myoma) was observed in the initial contrast-enhanced MRI, although it markedly decreased during 10-minute dynamic scanning and was measured to be 2.7\% in the post-therapy scan. NPV ratio less than 30\% has been shown to indicate sub-optimal treatment outcome in terms of symptom severity score~\citep{fennessy2007uterine}. The patient was admitted to surgical reintervention less than a month after the therapy.

\section*{Methods}

\subsection*{Perfusion analysis}

\begin{table}[t!]
  	\centering
  	\caption{Perfusion imaging parameters}
	\begin{tabular}{llll}
	\hline
	Parameter 				& Value 	& Parameter 			& Value							\\
	\hline
	Sequence 				& e-THRIVE 	& Repetition time 		& 2.93 ms						\\
	Fat saturation 			& SPAIR		& Echo time 			& 1.31 ms 						\\
	Imaging Plane 			& Axial 	& Temporal resolution 	& 4.18 ms						\\
							&			& Flip angle 			& 10$^\circ$		\\
							&			& Section thickness 	& 3 mm		\\
							&			& Field of view 		& 270 $\times$ 350 $\times$ 120 mm$^{3}$\\
	\hline
	\end{tabular}
  	\label{tab:perfusion}
  	\vspace{-0.5cm}
\end{table}

Pre-treatment MRI scans of the patient were performed four months before the HIFU treatment using a 3T MRI system (Ingenia, Philips Healthcare, Best, the Netherlands) with a 32-channel array torso coil. 

The MRI protocol comprised of dynamic contrast enhanced (DCE) T1-weighted sequence with parameters shown in the Table~\ref{tab:perfusion}. A single dose (0.2~mL/kg) of contrast agent (Dotarem, Guebert, Roissy, France) was injected at constant rate followed by 10~mL saline flush after the acquisition of five dynamic scans.

DCE-MRI data were analysed with NordicICE perfusion analysis software (v. 4.1.1, NordicNeuroLab AS, Bergen, Norway). First, the signal was converted to change in 1/T1 relaxation rate. Baseline T1 mapping was performed with inversion recovery technique. Arterial input function was determined from iliac artery by placing a circular region of interest (ROI) on the artery. The blood flow (i.e., the perfusion rate) was quantified by T1 perfusion deconvolution arithmetic with a first pass of arterial input function curve. The perfusion values were obtained from the entire fibroid and myometrium by averaging over three subsequent ROIs.

\subsection*{Ultrasound simulations}

\begin{figure}[b!]
    \centering
    \subfigure[]
    {
        \includegraphics[width=0.45\columnwidth]{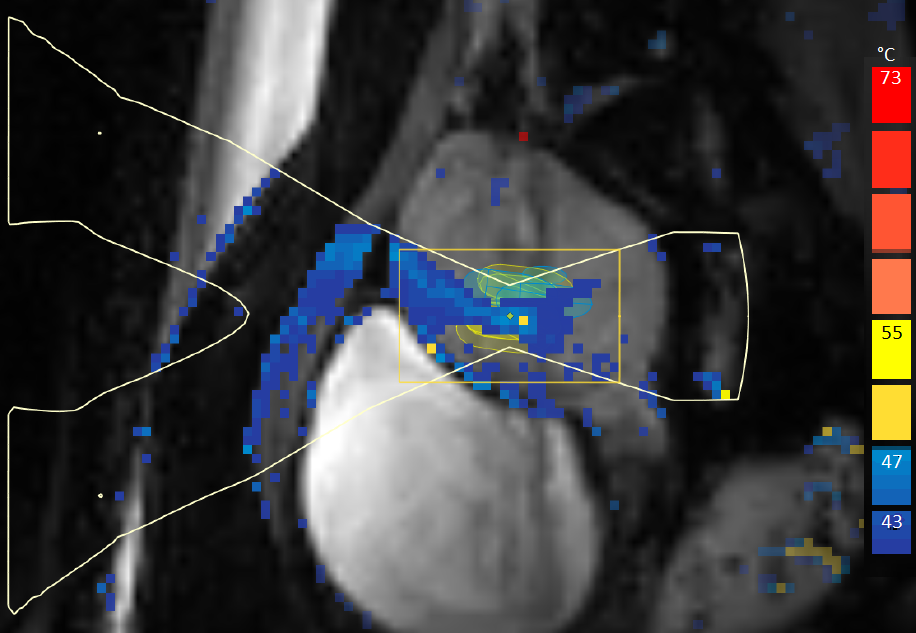}
    }
    \subfigure[]
    {
        \includegraphics[width=0.45\columnwidth]{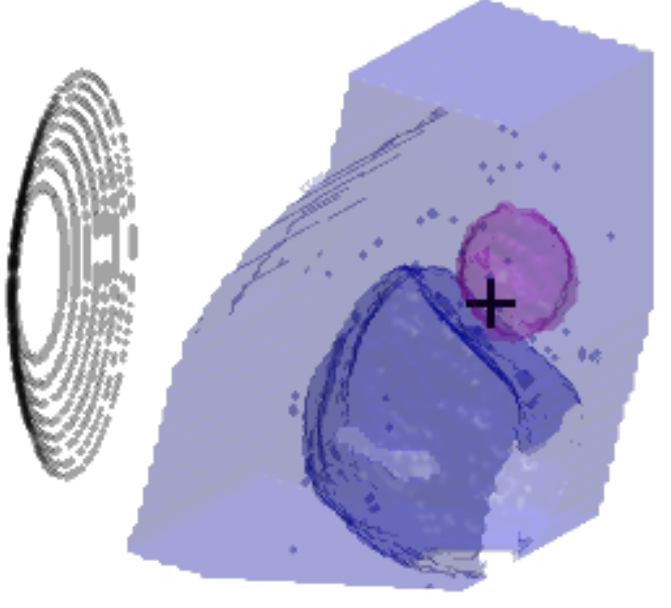}
    }
    \caption{(a) MRI temperature map acquired during HIFU therapy shows poor heating in the myoma when the transducer was tilted. (b) The three-dimensional simulation grid segmented using the MR images. The transducer was rotated around the left-right axis between 0 and $-$20 degrees with 5-degree steps while keeping the focal point  position inside the myoma (black cross) constant.}
    \label{fig:simulation_geometry}
\end{figure}

Three-dimensional ultrasound simulations were conducted using the patient MRI data acquired during the HIFU treatment (see Fig.~\ref{fig:simulation_geometry}(a)). The image data were segmented into water/bladder, bone, muscle, fat and myoma tissues (see Table~\ref{tab:acoustic_parameters} for acoustic properties~\citep{mast2000empirical, keshavarzi2001attenuation, hasgall2015database}). The simulations were conducted on the HIFU therapy system which had an annular transmitting surface of outer diameter 12.0~cm, inner hole diameter 4.0~cm, sonication frequency 1.2~MHz and focal length 14.0~cm. The simulations were linear with the computational grid supporting frequencies up to 2.0~MHz and they were carried out using the parallelised C++ version of the open source k-Wave Toolbox~\citep{jaros2016full}.

An illustration of the simulation geometry is presented in Fig.~\ref{fig:simulation_geometry}(b). The transducer was positioned so that the geometric focal point of the transducer was located in the bottom part of the myoma, which corresponded to the actual treatment location. The transducer was rotated at different angles around the left-right axis between 0 and $-$20 degrees relative to the target location inside the myoma while keeping the focal point position constant. Focal point deformation, maximum pressure values and focal shifts in axial, lateral and elevation directions were characterised and the results were then compared to the actual treatment outcome of the same patient.

\begin{table}[t!]
  	\centering
  	\caption{Tissue acoustic properties for ultrasound simulations}
    \begin{tabular}{lccc}
    \hline
          				& Density   		& Sound speed   & Attenuation 				 	\\
          				& (kg/m$^{3}$) 		& (m/s) 		& (dB/MHz$^{1.1}$/cm)			\\
    \hline
    Water/Bladder 		& 1000  			& 1520  		& 0.00217 					 	\\
    Bone  				& 1908  			& 3515  		& 4.74    					 	\\
    Muscle 				& 1050  			& 1547  		& 1.09   					 	\\
    Fat   				& 950   			& 1478  		& 0.48  						\\
    Myoma				& 1105  			& 1611			& 0.90  						\\
    \hline
    \end{tabular}
  	\label{tab:acoustic_parameters}
\end{table}

\clearpage

\section*{Results}

\subsection*{Perfusion analysis}

\begin{figure}[b!]
    \centering
    \subfigure[]
    {
        \includegraphics[width=0.46\columnwidth]{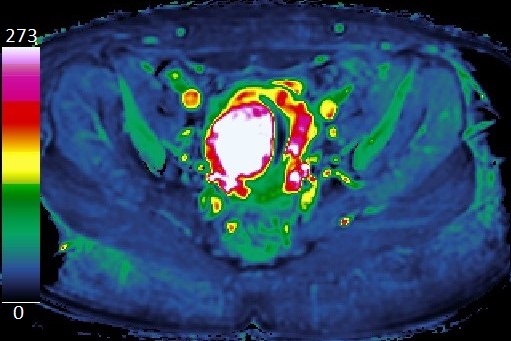}
    }
    \subfigure[]
    {
        \includegraphics[width=0.46\columnwidth]{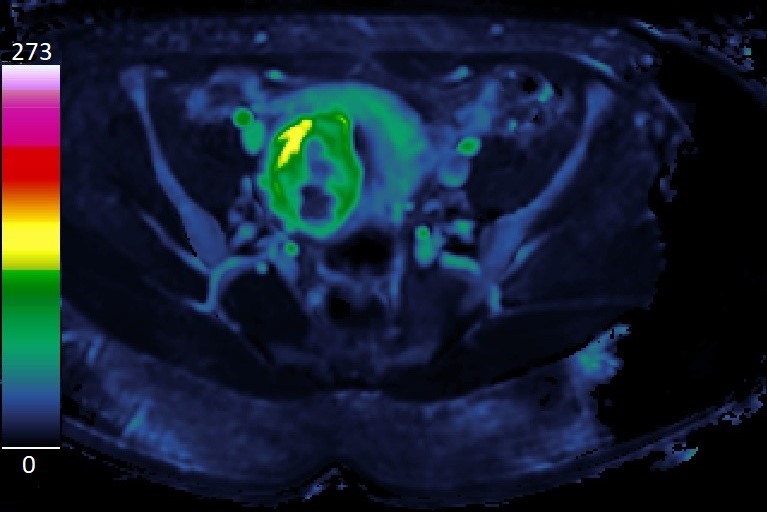}
    }
    \\
    \subfigure[]
    {
        \includegraphics[width=0.46\columnwidth]{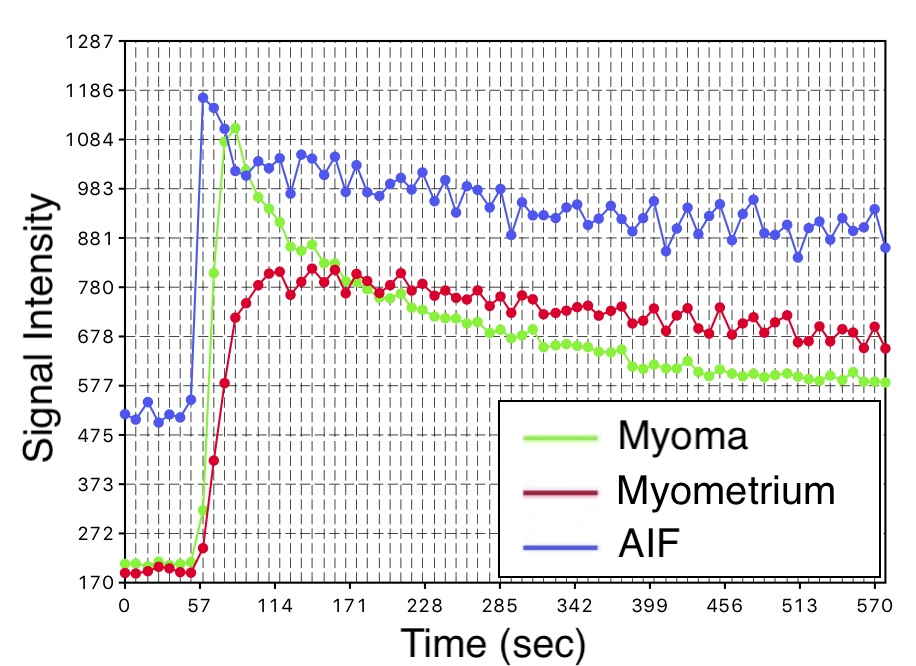}
    }
    \subfigure[]
    {
        \includegraphics[width=0.46\columnwidth]{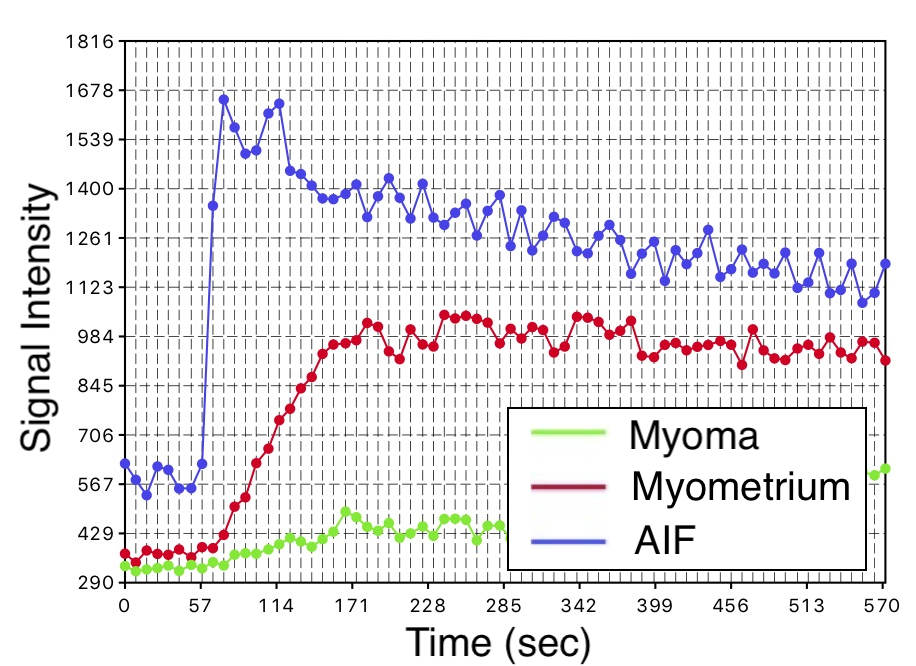}
    }
    \caption{(a)-(b) Axial DCE-MRI perfusion maps before and after the HIFU therapy, respectively, and (c)-(d) the corresponding signal intensity curves for the arterial input function, myometrium and myoma. The perfusion rate of the fibroid was markedly decreased after the therapy.}
    \label{fig:perfusion}
\end{figure}

Fig.~\ref{fig:perfusion}(a) shows the pre-therapy perfusion map of the axial plane acquired using DCE-MRI where the enhanced blood flow in the area of the uterine fibroid is clearly visible. The corresponding perfusion values were 301.0 $\pm$ 25.6 and 233.8 $\pm$ 16.2~mL/100 g/min in the myoma and myometrium, respectively. For reference, the perfusion rate of the gluteus maximus was 30.1 $\pm$ 3.7~mL/100 g/min.

Fig.~\ref{fig:perfusion}(b) shows the perfusion map after the therapy. The perfusion in the fibroid was markedly decreased despite of the rather small NPV. In the initial, arterial phase of the dynamic DCE-MRI NPV appeared quite decent, however, it markedly decreased during the 10-minute dynamic scanning and was measured only 2.7\% in the post therapy scan. When looking at the dynamic enhancement curves of the fibroid and normal myometrium before and after the treatment (see Fig.~\ref{fig:perfusion}(b) and (c)), it seems that this appearance was caused by the rather slow uptake of the contrast agent with markedly decreased arterial peak in the fibroid compared to the pre-treatment image. This could be explained, for example, by a transient vasospasm caused by the treatment.

\subsection*{Ultrasound simulations}

\begin{figure}[b!]
    \centering
    \subfigure[]
    {
        \includegraphics[height=3.9cm]{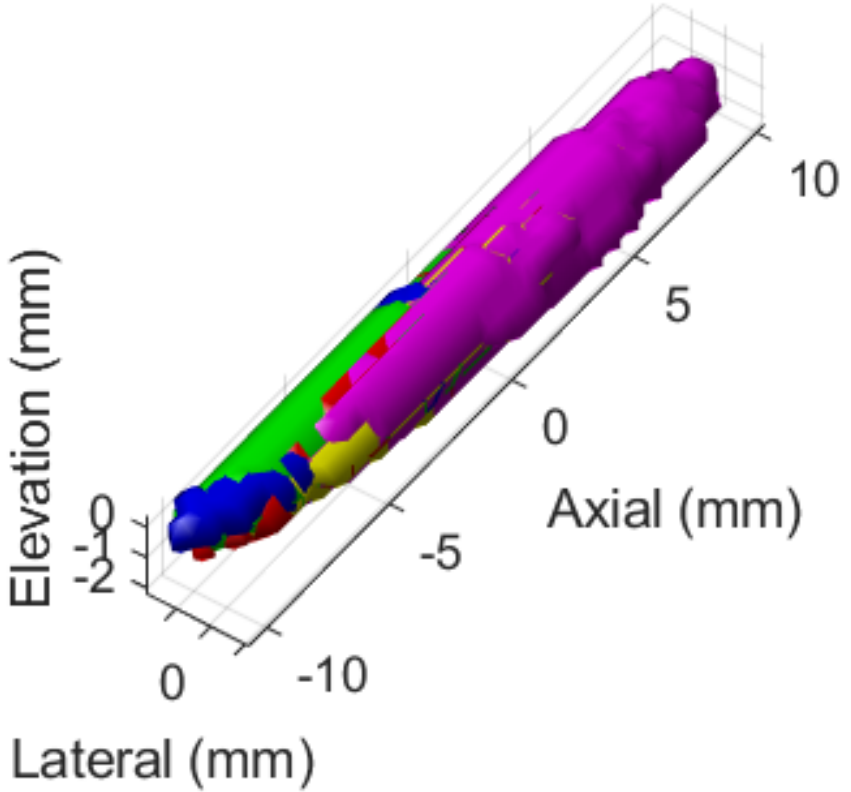}
    }
    \subfigure[]
    {
        \includegraphics[height=3.9cm]{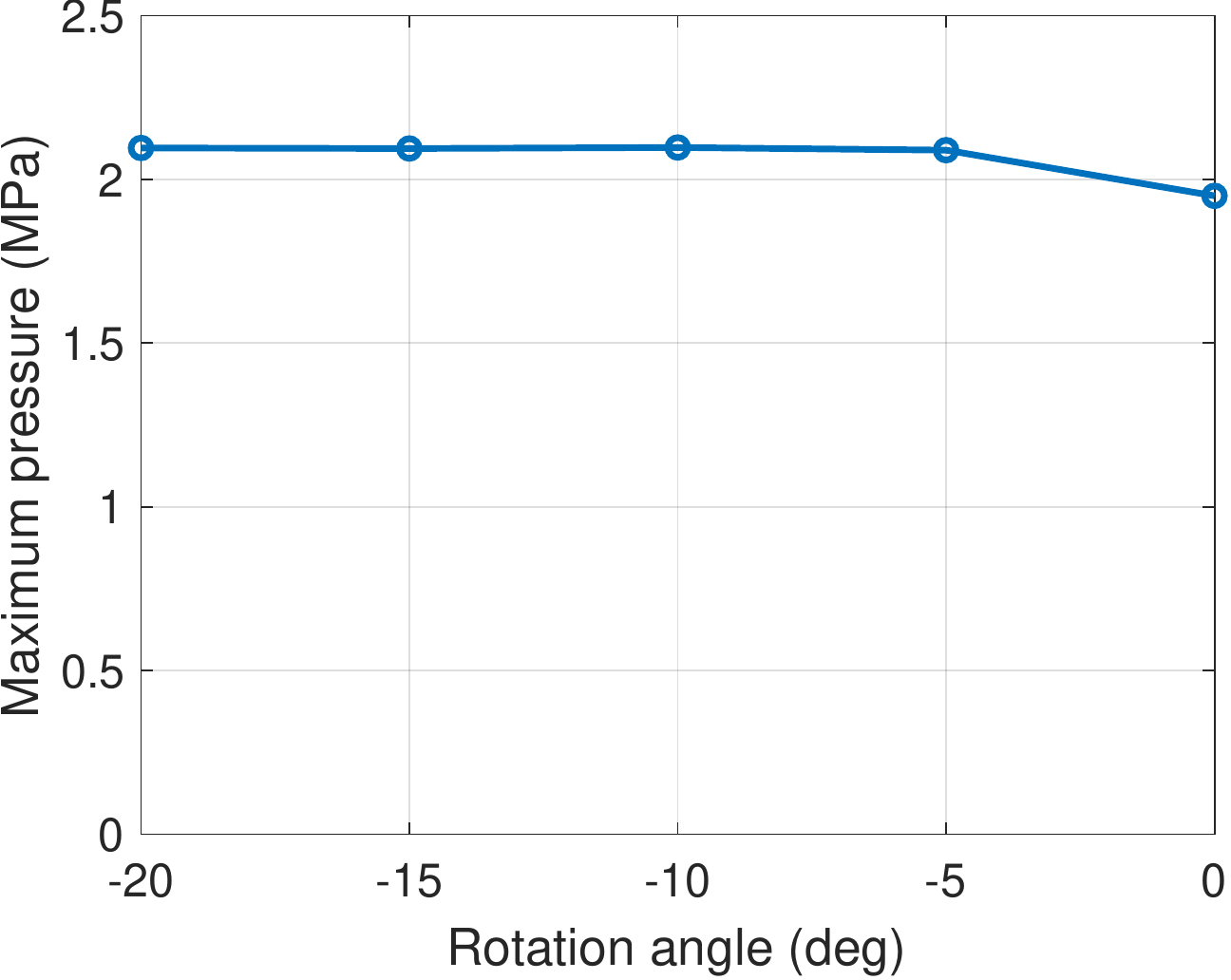}
    }
    \subfigure[]
    {
        \includegraphics[height=3.9cm]{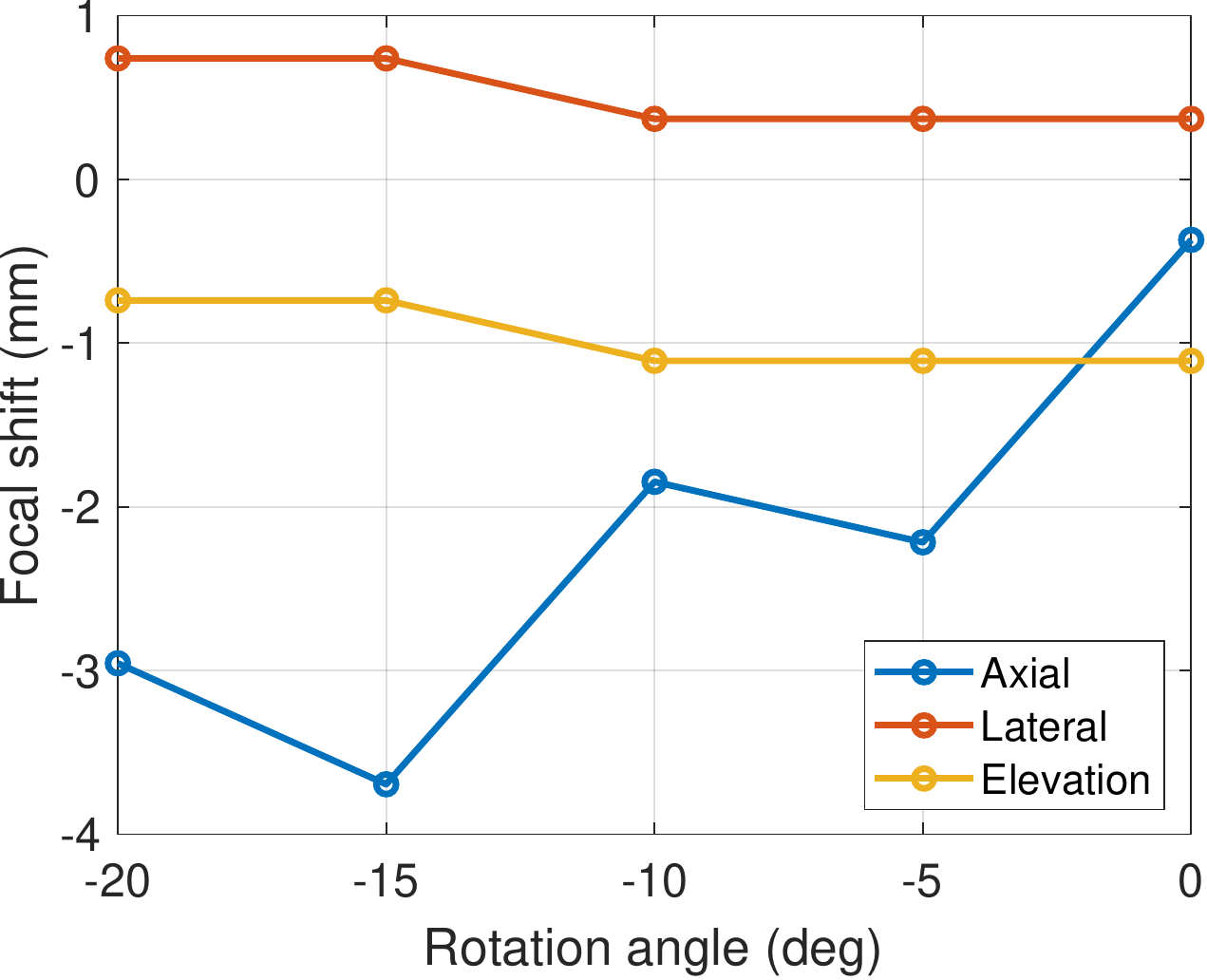}
    }
    \caption{(a) Colour-coded ultrasound focal points thresholded at $-$6~dB show the effect of transducer rotation on its shape and location (0$^{\circ}$ – magenta, $-$5$^{\circ}$ – yellow, $-$10$^{\circ}$ – red, $-$15$^{\circ}$ – green, $-$20$^{\circ}$ – blue). (b) The maximum pressure of the ultrasound field with respect to the transducer rotation angle. (c) The shifts in the focal point location (i.e., the maximum pressure) in axial, lateral and elevation directions with respect to the transducer rotation angle.}
    \label{fig:simulation_results}
\end{figure}

Changing the transducer rotation angle between 0 and $-$20~degrees did not have a large effect on the focal point shape where nearly ellipsoidal $-$6~dB focal region of the same size was observed in all the cases (see Fig.~\ref{fig:simulation_results}(a)). The focal volumes were approximately 18~mm in length and 3~mm in width and they had uniform isosurfaces. This suggests that the geometry of the ultrasound pathway through different tissue layers did not change considerably in order to deform or split the focal point into smaller subvolumes. This is partly because a large portion of the ultrasound pathway was inside the bladder which has negligible effect on ultrasound attenuation and dispersion.

Similarly, the maximum pressures inside the focal points did not change noticeably with the average maximum pressure of 2.07~MPa and standard deviation of $\pm$0.06~MPa (see Fig.~\ref{fig:simulation_results}(b)). For reference, the maximum pressure in water with the same output power was 4.49~MPa. Since focal point deformation or splitting were not observed, the pressure distributions inside the focal volumes also remained relatively constant with the maximum pressure being in the centre of the focal volume. Only at 0~degrees of transducer rotation, a slight drop in the maximum pressure was observed. This is probably due to the fact that the portion of ultrasound waves travelling through tissue is the greatest in this case, which results in high energy loss due to attenuation. On the contrary, the portion of waves propagating through the bladder is the smallest. The highest acoustic pressure (and thus, the heating efficacy) is achieved when the majority of ultrasound waves propagate through the bladder.

The effect of rotation angle on the focal shift (i.e., the location of the maximum pressure) was more pronounced (see Fig.~\ref{fig:simulation_results}(c)). In the axial, lateral and elevation directions average focal shifts of $-$2.2 $\pm$ 1.3~mm, 0.5 $\pm$ 0.2~mm and $-$1.0 $\pm$ 0.2~mm were observed, respectively. The simulation geometry does not change in lateral direction due to the rotation which explains the negligible shift in this direction. In elevation direction, the rotation angles might not be extreme enough in order to induce larger shifts than observed in the simulations. The axial shift can be attributed to the change in the geometry which `pulls' the focus towards the transducer at large rotation angles due to refraction and phase velocity changes.

\section*{Discussion}

An analysis of the factors affecting the heating efficacy during the HIFU therapy was conducted. For the analysis, retrospective clinical data of a single patient treated at the hospital for the uterine fibroid was used. Furthermore, DCE-MRI data was used for the perfusion analysis and patient-specific ultrasound simulations were conducted in order to evaluate the changes in acoustic parameters.

Poor heating efficacy was observed during the HIFU therapy although the maximum power allowed by the system was used. Rotating the transducer diminished the heating effect even further. The end result was  suboptimal and the patient was thus admitted to a surgical reintervention less than a month after the treatment.

During the therapy, a wedged gel pad was used to manipulate the myoma location to an optimal position for the therapy. In theory, the asymmetric shape of the gel pad related to the symmetry axis of the transducer could generate phase shifts in the propagating ultrasound waves. This in turn would deform or split the focal point at the target location and diminish the heating effect~\citep{suomi2018full}. However, the usage of the wedged gel pad was not found to affect the focal point shape nor the acoustic parameters in the simulations.

Similarly, changing the orientation of the transducer during the HIFU treatment did not result in a considerable loss of therapeutic efficacy in terms of the ultrasound focal point shape and acoustic pressure. The maximum acoustic pressure was observed to stay approximately constant despite of the rotation angle of the transducer. Likewise, the focal volume maintained its ellipsoidal shape at different angles of transducer orientation. Only a slight axial shift was observed when the rotation angle of the transducer was increased. In practice, however, the observed shift would not affect the therapy region since the thermal ablation volume is typically multiple times larger compared to the ultrasound focal volume~\citep{suomi2018full}.

The perfusion analysis showed enhanced blood flow in the region of the myoma when compared to the surrounding myometrium. A high perfusion rate has a strong cooling effect which has been shown to be a significant predictor of treatment outcome~\citep{keserci2017role}. It has previously been reported that the perfusion of the uterine fibroids varies on average between 50-334~mL/100 g/min, which supports that the fibroid in this study had relatively high perfusion rate~\citep{wei2018predictive}. Furthermore, it has been shown that in the prostate the perfusion rate significantly increases during thermal exposure~\citep{van2002prostate}, which could be the case in the myoma as well. The cooling effect of perfusion could potentially be diminished by the administration of oxytocin before the treatment~\citep{lozinski2018oxytocin}.

\section*{Conclusions}

In the evaluation of the clinical treatment outcome, the changes in the acoustic parameters of the ultrasound focal point did not seem to be the major cause for the experienced poor heating efficacy. Other factors, such as local perfusion, ultrasound attenuation and the location of the myoma, likely played a more significant role in this case. However, these results might differ when using another patient data, which still requires further research.

\section*{Acknowledgements}

V.~S. acknowledges the support of the State Research Funding (ERVA), Hospital District of Southwest Finland, number K3007 and CSC - IT Center for Science, Finland, for computational resources. The authors acknowledge J.~Jaros and B.~Treeby for developing and optimising the performance of the simulation model.

\pagebreak

\bibliographystyle{UMB-elsarticle-harvbib}
\bibliography{arXiv_embc}


\end{document}